\documentclass[twocolumn,prb,showpacs,preprintnumbers,superscriptaddress,amsmath,amssymb,citeautoscript]{revtex4-1}
\usepackage{graphicx}
\usepackage{epstopdf}
\usepackage{dcolumn}
\usepackage{color}
\usepackage{bm}
\usepackage{multirow}
\usepackage{rotating}
\usepackage{array}

\newcommand{\bpm}{\begin{pmatrix}}
\newcommand{\epm}{\end{pmatrix}}
\newcommand{\bs}{\boldsymbol}
\newcommand{\be}{\begin{equation}}
\newcommand{\ee}{\end{equation}}
\newcommand{\beq}{\begin{eqnarray}}
\newcommand{\eeq}{\end{eqnarray}}

\begin{document}

\title{Majorana fermions in finite-size strips in the presence of orbital effects: what is the Majorana polarization telling us?}
\author{Selma Franca}
\affiliation{Institut de Physique Th\'eorique, Universit\'e Paris Saclay, CEA
CNRS, Orme des Merisiers, 91190 Gif-sur-Yvette Cedex, France}
\affiliation{IFW Dresden, Helmholtzstra\ss e 20, 01069 Dresden, Germany}
\author{Vardan Kaladzhyan}
\email{vardan.kaladzhyan@phystech.edu}
\affiliation{Institut de Physique Th\'eorique, Universit\'e Paris Saclay, CEA
CNRS, Orme des Merisiers, 91190 Gif-sur-Yvette Cedex, France}
\affiliation{Department of Physics, KTH Royal Institute of Technology, Stockholm, SE-106 91 Sweden}
\author{Cristina Bena}
\affiliation{Institut de Physique Th\'eorique, Universit\'e Paris Saclay, CEA
CNRS, Orme des Merisiers, 91190 Gif-sur-Yvette Cedex, France}

\date{\today}

\begin{abstract}
We study numerically the formation of Majorana bound states in a finite-size quasi-one-dimensional square-lattice strip with Rashba spin-orbit coupling, in the presence of a proximity-induced superconducting pairing and a magnetic field perpendicular to the strip. We take into account both the Zeeman and orbital effects of the field. First, using the Majorana polarization, we demonstrate that such a system can host more than one pair of Majorana quasiparticles. We construct the corresponding topological phase diagram and we conclude that the topological regions are extended in the presence of orbital effects, however the gap protecting the topological states is reduced. 
\end{abstract}

\maketitle

\section{Introduction} 

After the Kitaev's groundbreaking article on realizing Majorana fermions in quantum wires\cite{Kitaev2001}, various condensed matter setups of different dimensionality were suggested as possible platforms hosting them. While 2D proposals appeared earlier\cite{Read2000,Fu2008}, 1D setups that were suggested theoretically to exhibit Majorana bound states\cite{Oreg2010,Lutchyn2010} motivated numerous promising experiments \cite{Lutchyn2018}. A typical 1D configuration showing signatures of Majorana bound states (MBS) is a semiconducting wire with proximity-induced superconducting pairing, spin-orbit coupling and a Zeeman field \cite{Oreg2010,Lutchyn2010}. These three ingredients are mimicking a spinless p-wave superconductor thus providing a condensed matter emulation of the Kitaev model. 

In many previous works the orbital effects associated with the magnetic field were not taken into account. This simplification of the problem to the only three aforementioned ingredients is possible in certain situations, namely, if the wire is very thin and if the magnetic field is parallel to the axis of the wire. Nevertheless, these conditions can be easily violated in experimental setups where, first, the Zeeman field contribution can be much smaller than that of the orbital field\cite{Mourik2012}, and second, some misalignment of the applied magnetic fields is unavoidable. The corresponding orbital effects have been reported to be destructive for the Majorana states when cylindrical and hexagonal wires with different superconducting coatings were considered \cite{Nijholt2016}, or in planar nanowires under certain conditions\cite{Lim2012}. However, it was shown that in other situations orbital effects do not necessarily undermine the formation of MBS\cite{Osca2015,Kiczek2017}, and moreover, can favor their appearance by shortening the corresponding coherence length \cite{Dmytruk2018}.

In this work we consider the simple case of a planar quasi-one-dimensional system (a finite-size strip) with proximity-induced s-wave superconducting order and Rashba spin-orbit coupling, subject to a perpendicular magnetic field, for which it was shown that Majorana states may also arise \cite{Sedlmayr2016}. The perpendicular component of the magnetic field will produce orbital effects that may become relevant; some of these effects have been touched upon in Refs.~\onlinecite{Osca2015,Lim2012}. We revisit this setup using the new tool introduced in Refs.~\onlinecite{Sticlet2012,Sedlmayr2015b}, the Majorana polarization, which allows us to draw some new and remarkable conclusions.  

Thus we first show that in the presence of orbital effects such a setup may support \textit{multiple} Majorana modes. We present also the spatial structure of the Majorana states and point out drastic differences between different Majorana pairs. While it was demonstrated that due to orbital effects a cylindrical geometry may support multiple Majorana modes\cite{Lim2013}, no situations with multiple modes have been previously identified in a planar wire geometry\cite{Osca2015}. Moreover, we show that the formation of multiple Majorana pairs stems from the orbital-induced spatial variation of the effective superconducting gap that can split the strip into different regions with topological and trivial characters.
 
Second, we construct the corresponding phase diagrams as a function of the chemical potential and the magnetic field.  We find that the topological phase in the presence of the orbital effects is greatly extended with respect to that obtained in their absence, however the topological gap is strongly reduced by the orbital effects.   

The model Hamiltonian for our system is presented in Section II, the corresponding methods are described in Section III, the results and a discussion thereof are presented in Section IV and we conclude in Section V.

\section{Model Hamiltonian}

We start with a simple low-energy model describing an $s$-wave superconductor with Rashba spin-orbit coupling. The corresponding Hamiltonian can be written as: 
\begin{equation}
H=\int \frac{d\bs{p}}{(2\pi\hbar)^2}\;\Psi_{\bs{p}}^{\dag} \mathcal{H} \Psi_{\bs{p}}
\end{equation}  
where $\Psi^\dag_{\bs{p}}=(\psi_{\bs{p}\uparrow}^{\dag},\psi_{\bs{p}\downarrow}^{\dag},\psi_{-\bs{p}\downarrow},-\psi_{-\bs{p}\uparrow})$ is the Nambu basis, and
\begin{equation} \label{initialH}
\mathcal{H}_0=\bigg(\frac{\bs{p}^2}{2m}-\mu'\bigg)\tau_z-\Delta\tau_x+\alpha (p_y \sigma_x -  p_x \sigma_y) \tau_z,
\end{equation}
where $m$ is the effective mass of quasiparticles, $\mu'$ denotes the chemical potential, $\alpha$ the value of the Rashba spin-orbit coupling and $\Delta$ is the superconducting pairing. Pauli matrices $\bs{\sigma} \equiv (\sigma_x,\;\sigma_y,\;\sigma_z)$ and $\bs{\tau} \equiv (\tau_x,\;\tau_y,\;\tau_z)$ act in spin and particle-hole subspaces correspondingly. \\

In order to introduce an out-of-plane magnetic field $\bs{b} \equiv b \,\hat{n}_{z}$ into the model above we add the Zeeman energy $V_Z\sigma_z $ into the Hamiltonian in Eq.~(\ref{initialH}) and we make the Peierls substitution $\bs{p} \to \bs{\Pi} = \bs{p} + e \bold{A} \tau_z$ to take into account the orbital effects of the field.
For the latter we use the symmetric gauge in the following form: $\bold{A}=\frac{1}{2}(-b y,b x,0)$.  The relation between the Zeeman and the orbital field is given by $V_Z=g \mu_B b/2$ where $g$ is the $g$-factor of the material, and $\mu_B = e \hbar/ 2m_e$ is the Bohr magneton. All physical quantities characterizing this model, including the energy spectrum, should be independent of the choice of gauge. Within the symmetric gauge, Eq.~(\ref{initialH}) becomes: 
\begin{multline} \label{Hsymgauge}
\mathcal{H}=\bigg[\frac{(p_x - \frac{eby}{2} \tau_z)^2}{2m}+\frac{(p_y + \frac{ebx}{2} \tau_z)^2}{2m}-\mu'\bigg]\tau_z -\Delta\tau_x \\ +V_Z\sigma_z+\alpha\bigg[ \bigg(p_y + \frac{ebx}{2} \tau_z\bigg) \sigma_x  - \bigg(p_x - \frac{eby}{2}\tau_z\bigg) \sigma_y \bigg]\tau_z
\end{multline}

To diagonalize numerically the Hamiltonian above, we discretize it on a square lattice\cite{Governale1998,Janecek2008}. The problem of gauge invariance on the discrete lattice in the presence of superconductivity is discussed in more detail in Appendix A, as well as in Refs.~[\onlinecite{Governale1998}], [\onlinecite{Janecek2008}] and [\onlinecite{Osca2015}]. We leave the full derivation of the tight-binding form of the Hamiltonian (\ref{Hsymgauge}) in the symmetric gauge to Appendix A. Here we only note that after the Peierls substitution a given wavefunction acquires an additional phase factor: $\psi'=e^{-i \frac{e}{\hbar} \int{\bold{A} \bold{dl}}\; \tau_z} \psi$.  Furthermore, $\Pi_x\psi$ can be rewritten as: 
\begin{equation}
\left( -i \hbar \partial_x + e A_x\tau_z \right) \psi = e^{i \frac{e}{2\hbar} b x y \tau_z} \left(-i \hbar \partial_x \right)e^{-i \frac{e}{2\hbar} b x y\tau_z } \psi
\label{gaugephase}
\end{equation}
Note that an analogous expression can be derived for $\Pi_y \psi$. The final form of the tight-binding Hamiltonian obtained using the transformation introduced in Eq.~(\ref{gaugephase}) is given by:
\begin{eqnarray}
\nonumber \mathcal{H}_{\mathrm{TB}} = \sum\limits_{\bold{r}} \Bigg\{ \Psi_{\bold{r}}^\dag \Big[ -\mu \tau_z -\Delta \tau_x + V_Z \sigma_z \Big] \Psi_{\bold{r}} \phantom{aaa}\\
\nonumber +\Psi_{\bold{r}}^\dag \left[ \left( t + i \frac{\hbar \alpha}{2a} \sigma_y\right) \otimes e^{-i \frac{e b a}{2\hbar} y \tau_z}\tau_z \right] \Psi_{\bold{r}+\bold{x}}\\
+ \Psi_{\bold{r}}^\dag \left[ \left( t - i \frac{\hbar \alpha}{2a} \sigma_x\right) \otimes e^{i \frac{e b a}{2\hbar} x \tau_z}\tau_z \right] \Psi_{\bold{r}+\bold{y}} + \mathrm{H.c.} \Bigg\}
	\label{TBHamiltonian}
\end{eqnarray}
Here $\bold{r} \equiv (x,\,y),\; t \equiv -\frac{\hbar^2}{2ma^2}$, where $a$ denotes the lattice constant. The vectors $\bold{x}$ and $\bold{y}$  represent the unit vectors in the corresponding directions. To be consistent with conventions in previous works\cite{Sedlmayr2016,Kaladzhyan2017b}, in the onsite term we have introduced a shift of the chemical potential $\mu'$ by $4t$, $\mu \equiv \mu' - 4t$. Since it is easier to work with dimensionless quantities we introduce $\tilde{\mathcal{H}}_{\mathrm{TB}} =\mathcal{H}_{\mathrm{TB}}/t$:
\begin{eqnarray}
\nonumber \tilde{\mathcal{H}}_{\mathrm{TB}} = \sum\limits_{\bold{r}} \Bigg\{ \Psi_{\bold{r}}^\dag \Big[ -\tilde{\mu}\tau_z -\tilde{\Delta} \tau_x + \tilde{V}_Z \sigma_z \Big] \Psi_{\bold{r}} \phantom{aaa}\\
\nonumber +\Psi_{\bold{r}}^\dag \left[ \left( \mathbb{I}_2 - i \tilde{\alpha} \sigma_y\right) \otimes e^{i \frac{m_e}{g m}\tilde{V}_Z y \tau_z}\tau_z \right] \Psi_{\bold{r}+\bold{x}}\\
+ \Psi_{\bold{r}}^\dag \left[ \left( \mathbb{I}_2 + i \tilde{\alpha} \sigma_x\right) \otimes e^{-i \frac{m_e}{g m}\tilde{V}_Z x \tau_z}\tau_z \right] \negthickspace \Psi_{\bold{r}+\bold{y}} + \mathrm{H.c.} \Bigg\}
	\label{TBHamiltonianUnitsOft}
\end{eqnarray}
where $m_e$ stands for the rest mass of an electron, $\tilde{\alpha} \equiv m a \cdot \alpha / \hbar$ is the dimensionless spin-orbit coupling amplitude, and all energies ($\tilde{\Delta}$, $\tilde{V}_Z$, $\tilde{\mu}$) are now dimensionless and expressed in units of $t$, while $x$ and $y$ are expressed in units of $a$.

\section{Methods}
In this work fully-open systems are considered, hence one cannot diagonalize the Hamiltonian using Fourier transforms. Moreover, in the presence of orbital fields the calculation of bulk topological invariants cannot be done straightforwardly. Therefore, we resort to numerical methods for solving this problem, in particular the use of the MatQ code \cite {MatQ}. 

There are several useful methods of visualizing MBS in such systems. First, we calculate the spectrum and look for the zero-energy states. We also study the local density of states (LDOS) defined in real space as:
\begin{equation}
\rho(\bold{r},\epsilon)=\sum_n  \left\langle \psi_n(\bold{r})\left\vert \frac{\tau_0+\tau_z}{2} \right\vert \psi_n(\bold{r})  \right\rangle \delta(\epsilon_n-\epsilon)
\end{equation}
The expression above takes into account the LDOS of electrons and not of holes, since the two are not \textit{simultaneously} accessible in experimental setups. 

The most useful information is, however, recovered using the Majorana Polarization \cite{Sticlet2012,Sedlmayr2015a,Sedlmayr2015b,
Sedlmayr2016,Kaladzhyan2017a,Kaladzhyan2017b,
Sedlmayr2017}. While the LDOS is a scalar quantity, the Majorana polarization (MP) is a vector quantity since it measures the particle-hole overlap along with the relative phase between the electron and the hole components of the wavefunction. In order to classify a given state as a Majorana bound state it has to have a perfect overlap of the particle and hole components and thus have a MP equal to one. Same as the LDOS, the Majorana polarization is a property of the eigenfunctions of the Hamiltonian, so it cannot be extracted solely from the eigenvalues. The MP can be interpreted as a measure of the Majorana density at each position. 

If we introduce the particle-hole operator $\hat{C} = \sigma_y \tau_y \cdot \mathcal{K}$ ($\mathcal{K}$ stands for complex conjugation operator), we have
$ \hat{C} \gamma \equiv \hat{C} (c+c^{\dag})=c^{\dag}+c \equiv \gamma$. Thus, the eigenvalue of $\hat{C}$ for a MBS should be equal to one, and more precisely, for a spatial distribution \cite{Sedlmayr2016}: 
\begin{equation}
C=\frac{\sum\limits_{\bold{r}} \left\vert \left\langle\Psi(\bold{r})\left\vert \hat{C}\bold{r} \right\vert \Psi(\bold{r})\right\rangle \right\vert}{\sum\limits_{\bold{r}} \left\vert \langle\Psi(\bold{r})\left\vert \bold{r} \right\vert \Psi(\bold{r}) \rangle \right\vert} = 1
\end{equation}
The summation usually runs along half of the system. Unlike calculating energies or finding the LDOS, the MP gives a necessary and sufficient criterion for determining whether a given state is a MBS or not, thus being a versatile and indispensible tool for studying Majorana physics.

\section{Results}

\subsection{Quasi-one-dimensional strip}
\begin{figure}[h!]
	\begin{center}
	\includegraphics[width=0.49\columnwidth]{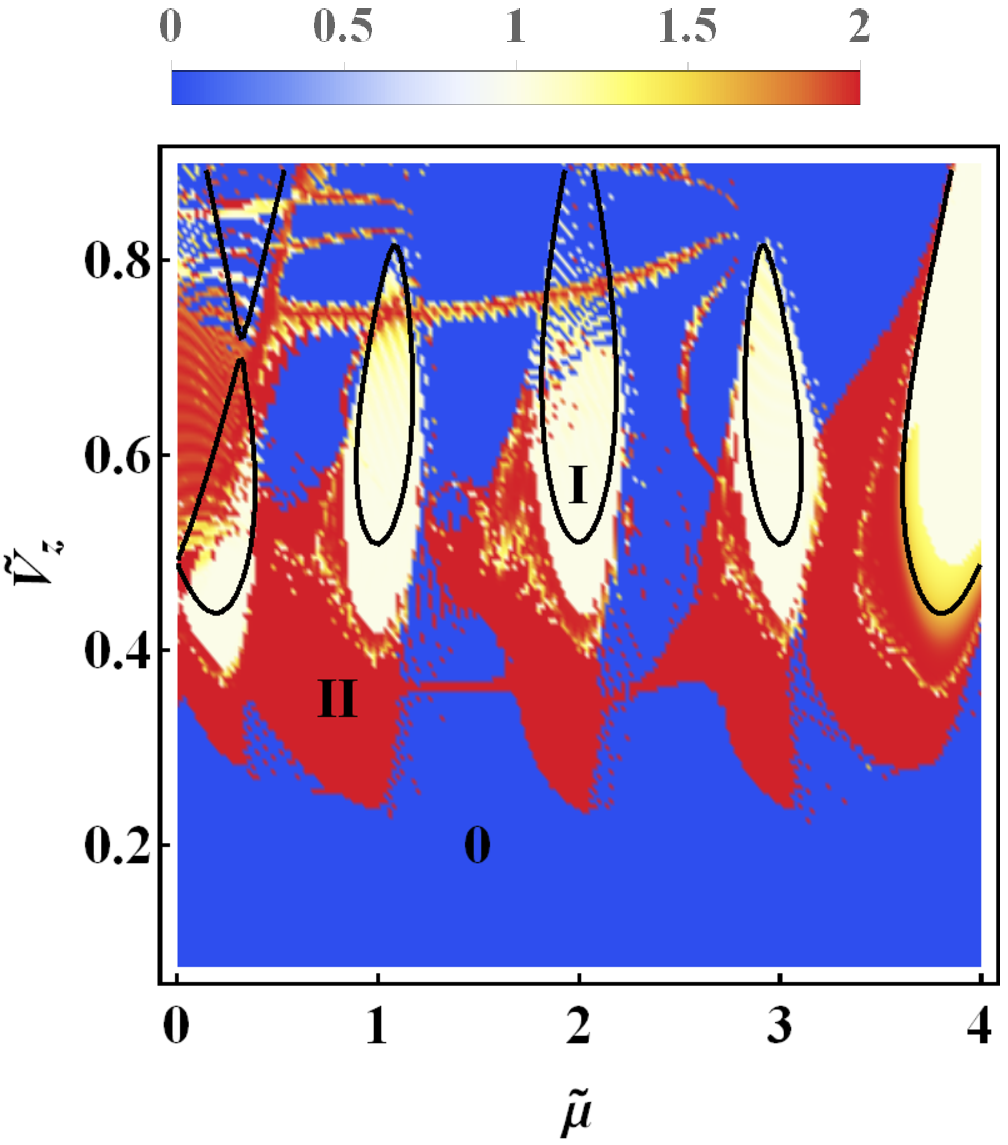}
	\includegraphics[width=0.49\columnwidth]{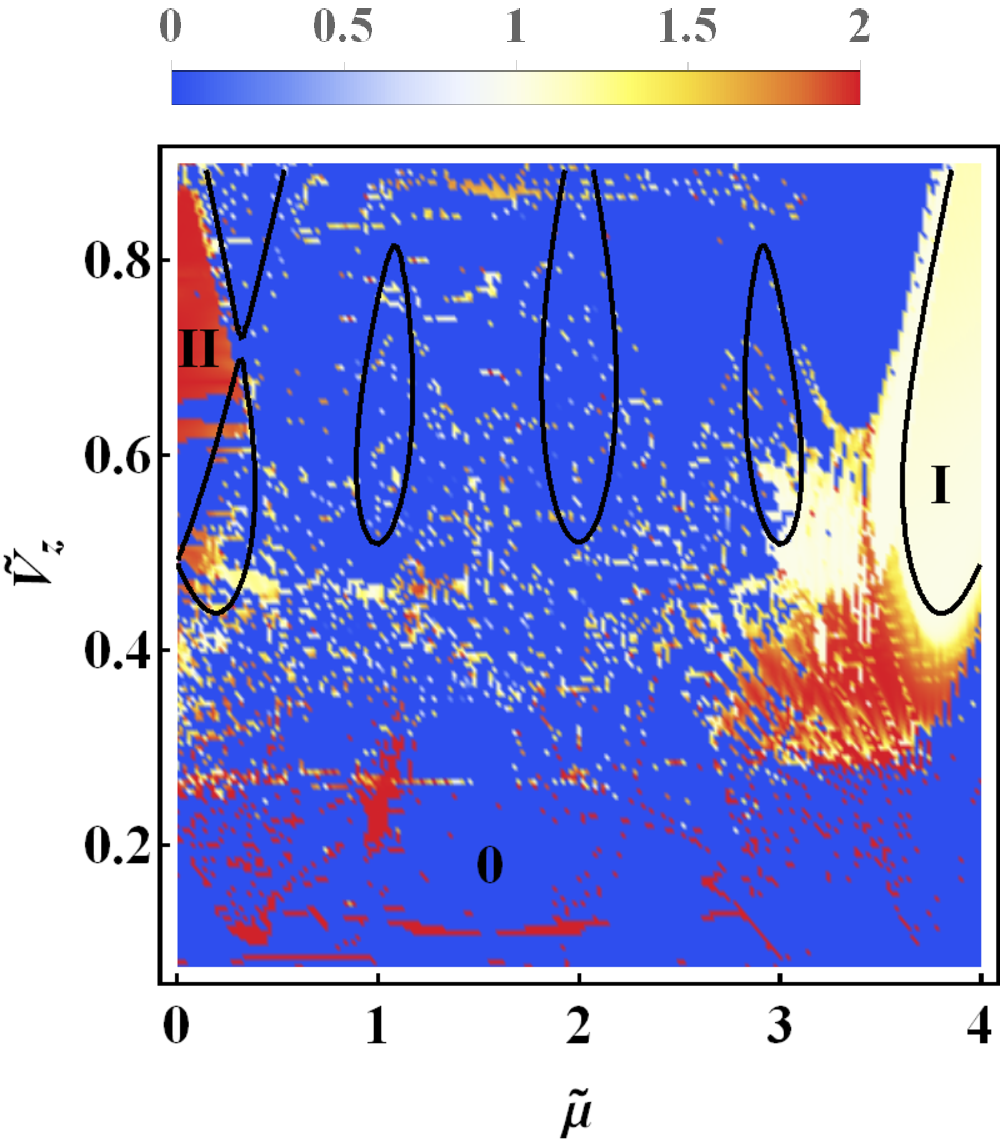}
	\end{center}
	\caption{Phase diagram in the presence of orbital effects. We plot the MP as a function of $\tilde{\mu}$ and $\tilde{V}_z$. The black lines indicate the boundary of the phase diagram when the orbital effects are ignored. The g-factor $g^*$ is taken to be 50 for the left panel and 10 for the right panel respectively. We consider $\tilde{\Delta}=0.4$, $\tilde{\alpha}=0.2$.}  
	\label{fig1}
\end{figure}

We first consider a quasi-one-dimensional square-lattice strip of dimensions $L_x \times L_y = 201 \times 5$ described by Eq.~(\ref{initialH}). Subject to an out-of-plane Zeeman field, this quasi-1D system can enter a topological phase hosting MBS in the absence of orbital effects \cite{Sedlmayr2016}. We calculate the sum of the MP for the four lowest energy states and over half the wire and in Fig.~\ref{fig1} we plot this quantity as a function of $\tilde{\mu}$ and the applied magnetic fields for two values of the effective $g$-factor, $g^*=g m/m_e$.
Note that for a large g-factor (i.e., when the Zeeman field dominates over the orbital effects), the phase diagram acquires extended topological regions at small magnetic fields even for $V_Z<\Delta$. These regions do not appear in the absence of orbital effects. Most remarkably, we note the apparition of regions in which two pairs of MBS arise. For smaller g-factors the orbital effects are becoming destructive, except for the region around $\tilde{\mu}=4$. 

Below we also plot the dependence of the MP with magnetic field and g-factor for two fixed chemical potentials. For $\tilde{\mu}=2$ (left panel) we see that for small orbital effects the topological phase is extended and the phase with two pairs of Majorana appears, but for larger  orbital  effects the topological phase is destroyed, in full accordance with previous works \cite{Lim2012,Lim2013,Serra2013,Osca2015,Nijholt2016,Kiczek2017}. However, the topological region close to $\tilde{\mu}=3.9$ (right panel) is preserved for quite large values of the orbital effect (small values of g). A full dependence on $\tilde{b}\equiv \tilde{V}_Z/g^*$ up to very large values corresponding to a quantum of flux per unit cell, as well as the special case of $\tilde{b}=\pi$ are discussed in Appendixes B and C.

\begin{figure}[h!]
	\begin{center}
	\includegraphics[width=0.49\columnwidth]{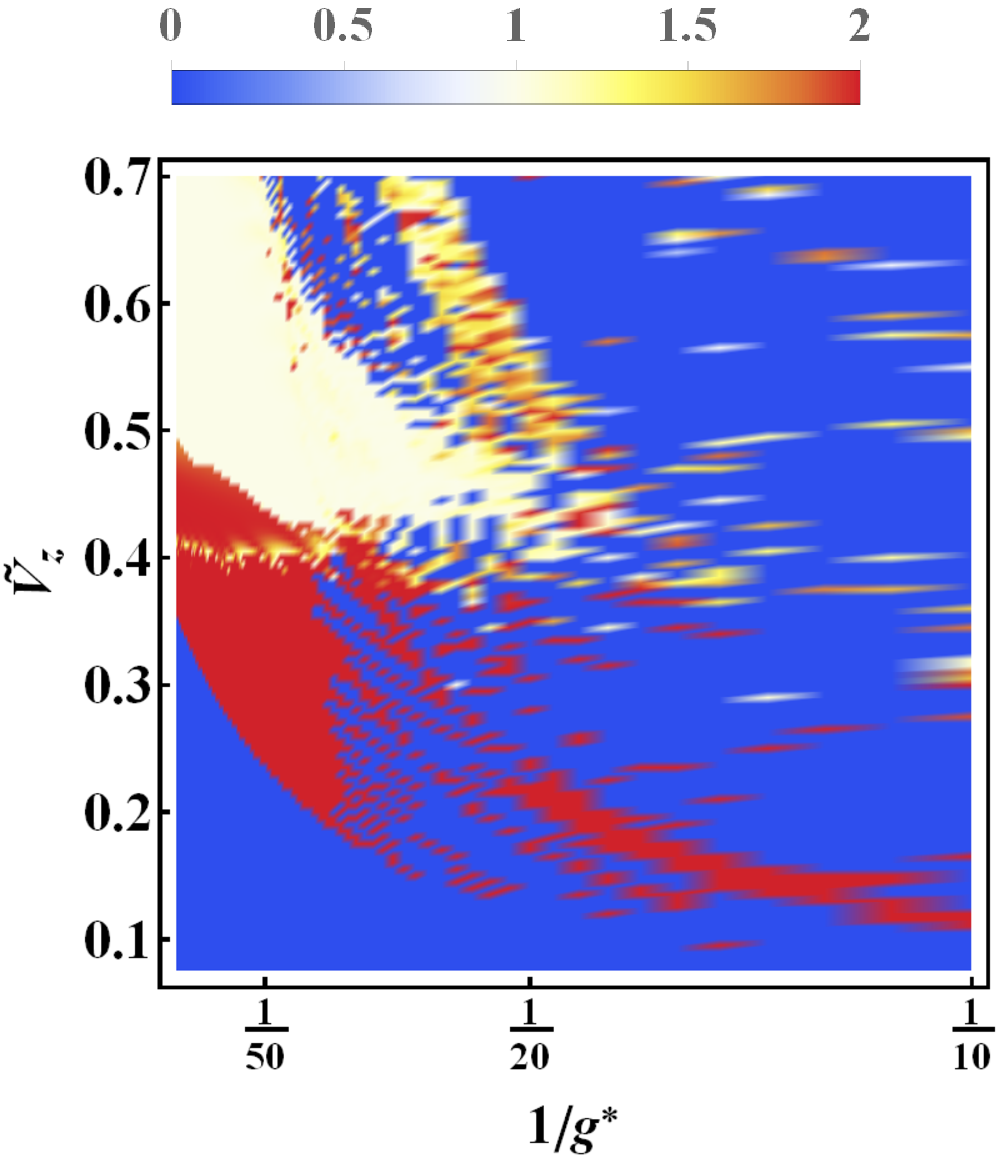}
	\includegraphics[width=0.49\columnwidth]{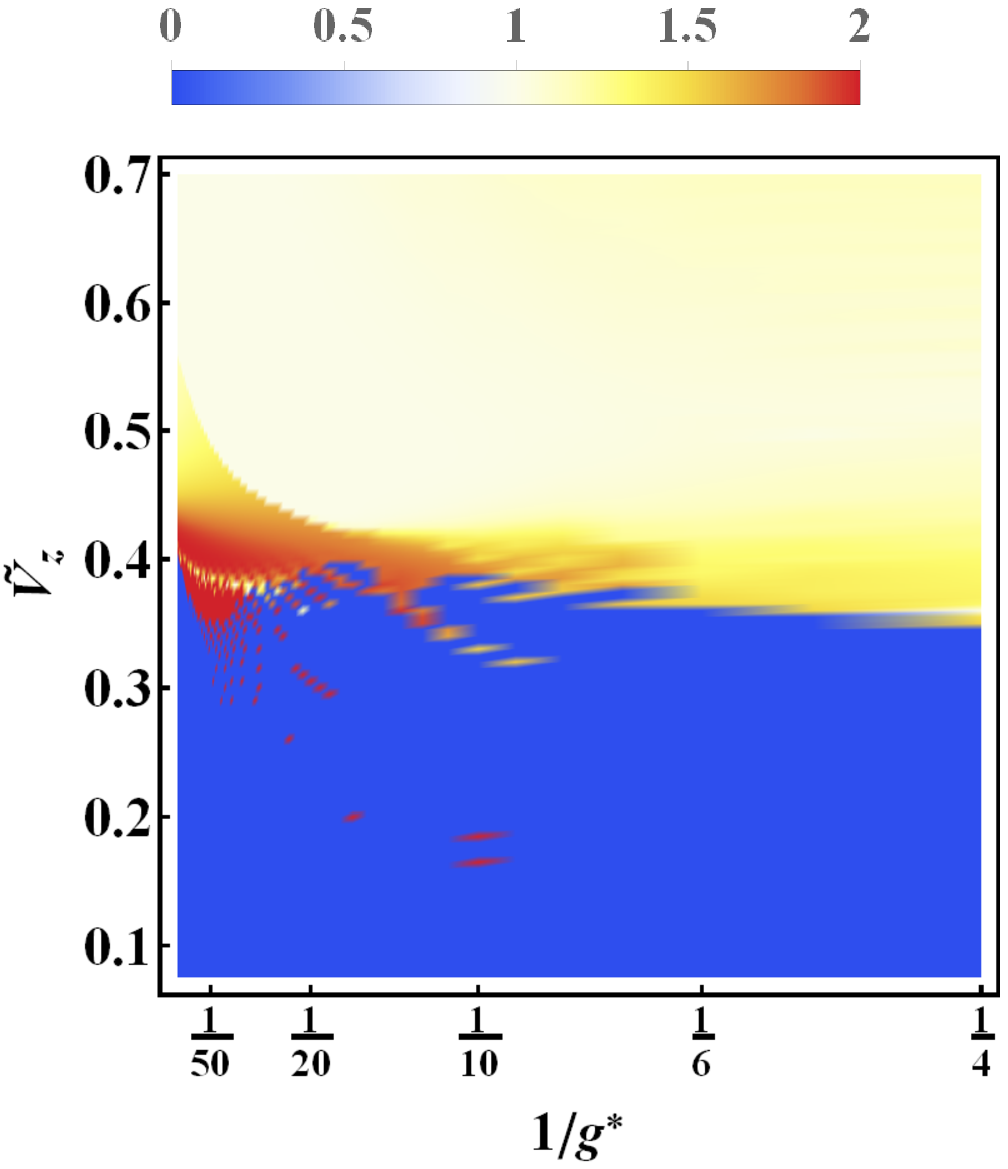}
	\end{center}
	\caption{The MP as a function of $\tilde{V}_z$ and the g-factor $g^*$. We set $\tilde{\Delta}=0.4$, $\tilde{\alpha}=0.2$, $\tilde{\mu}=2$ for the left panel and $\tilde{\mu}=3.9$ for the right one.}  
	\label{fig2}
\end{figure}

\begin{figure}[h]
	\begin{center}
	\includegraphics[width=1\columnwidth]{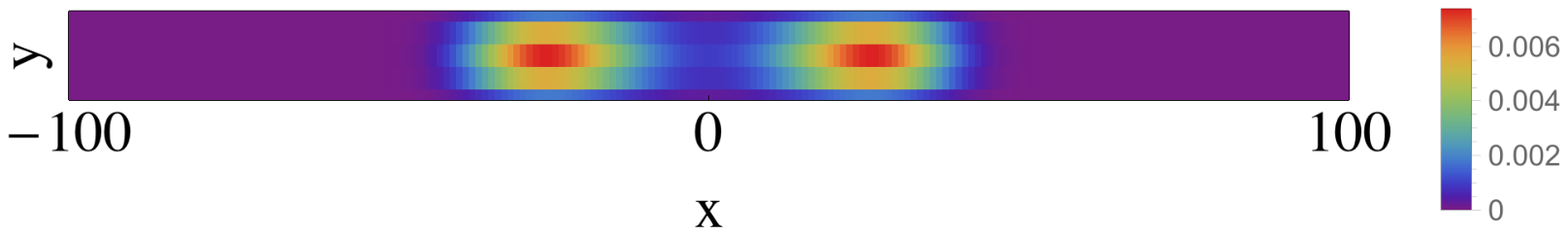}
	\includegraphics[width=1\columnwidth]{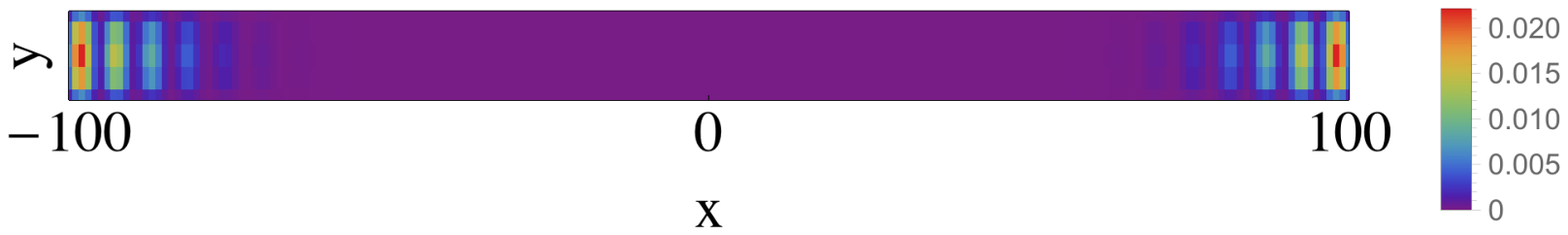}
	\includegraphics[width=1\columnwidth]{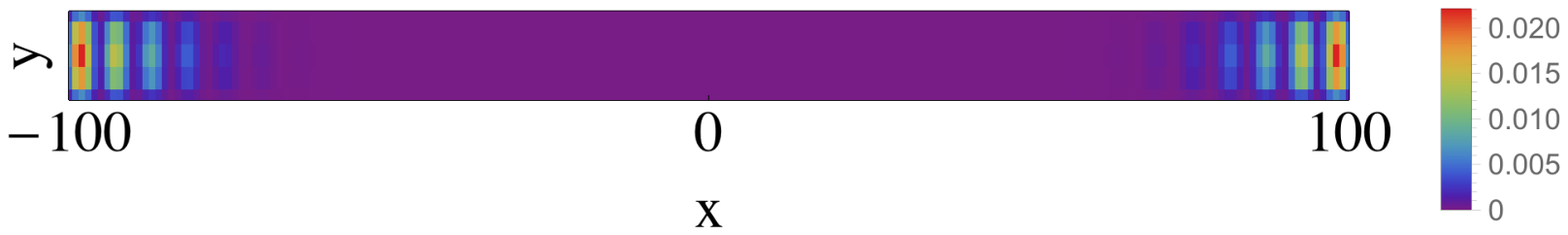}
	\includegraphics[width=1\columnwidth]{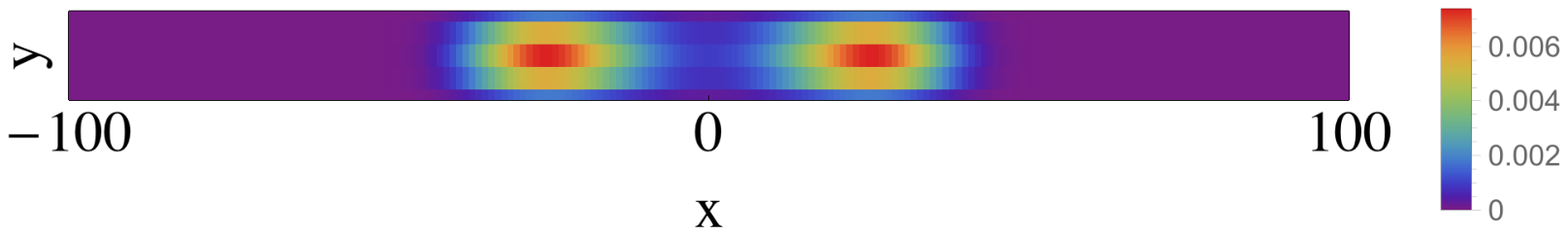}
	\end{center}
	\caption{The amplitude of the wavefunction for the four lowest energy states when these states are MBS. We consider $\tilde{\Delta}=0.4$, $\tilde{\alpha}=0.2$, $\tilde{\mu}=3.9$, $\tilde{V}_Z=0.4$, $\tilde{b}=0.4/50$ and $g^*=50$.}  
	\label{LDOSstrip}
\end{figure}

A typical phase diagram can be divided into various regions: 0 -- the trivial phase with $C=0$, denoted in blue, I -- the topological region with $C=1$ (i.e. one pair of Majorana states), denoted in white, and finally, II -- the topological region with $C=2$ and two pairs of Majorana states. 
Exemplary local densities of states for the latter case are shown in Fig.~\ref{LDOSstrip}.
The``traditional'' pair of states localized at the ends of the wire survive at low fields and large $g^*$'s for an extended parameter range compared to the case with no orbital effects (the white regions in Fig.~\ref{fig1} are much wider than the topological regions delimited by the black lines in the absence of orbital effects). The pair of states localized towards the center of the wire is more sensitive to disorder since they can hybridize more easily due to the shorter distance between them; moreover they do not form if the length of the wire is too small.

\begin{figure}[h!]
\begin{center}
\includegraphics[width=0.49\columnwidth]{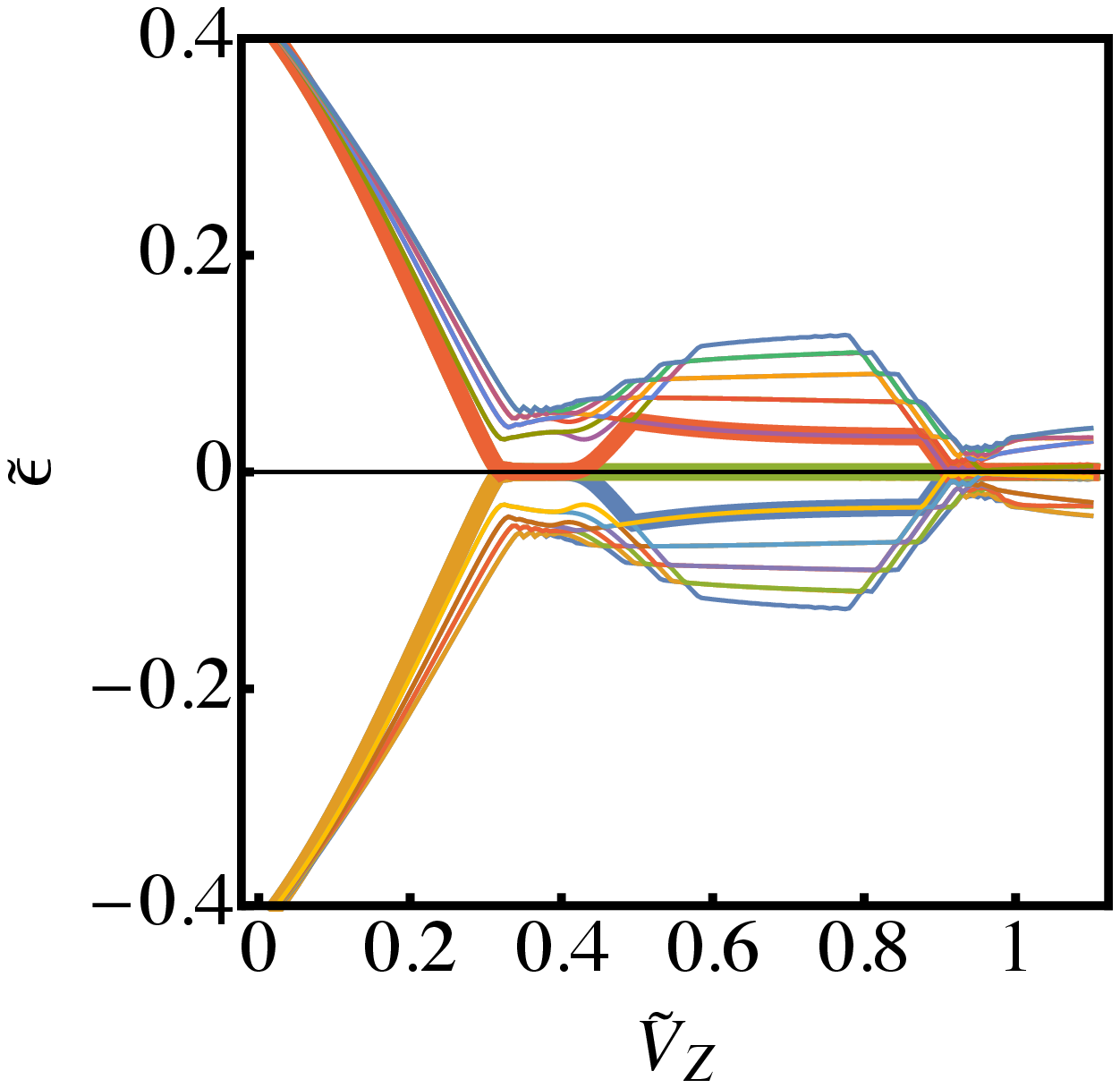}
\includegraphics[width=0.45\columnwidth]{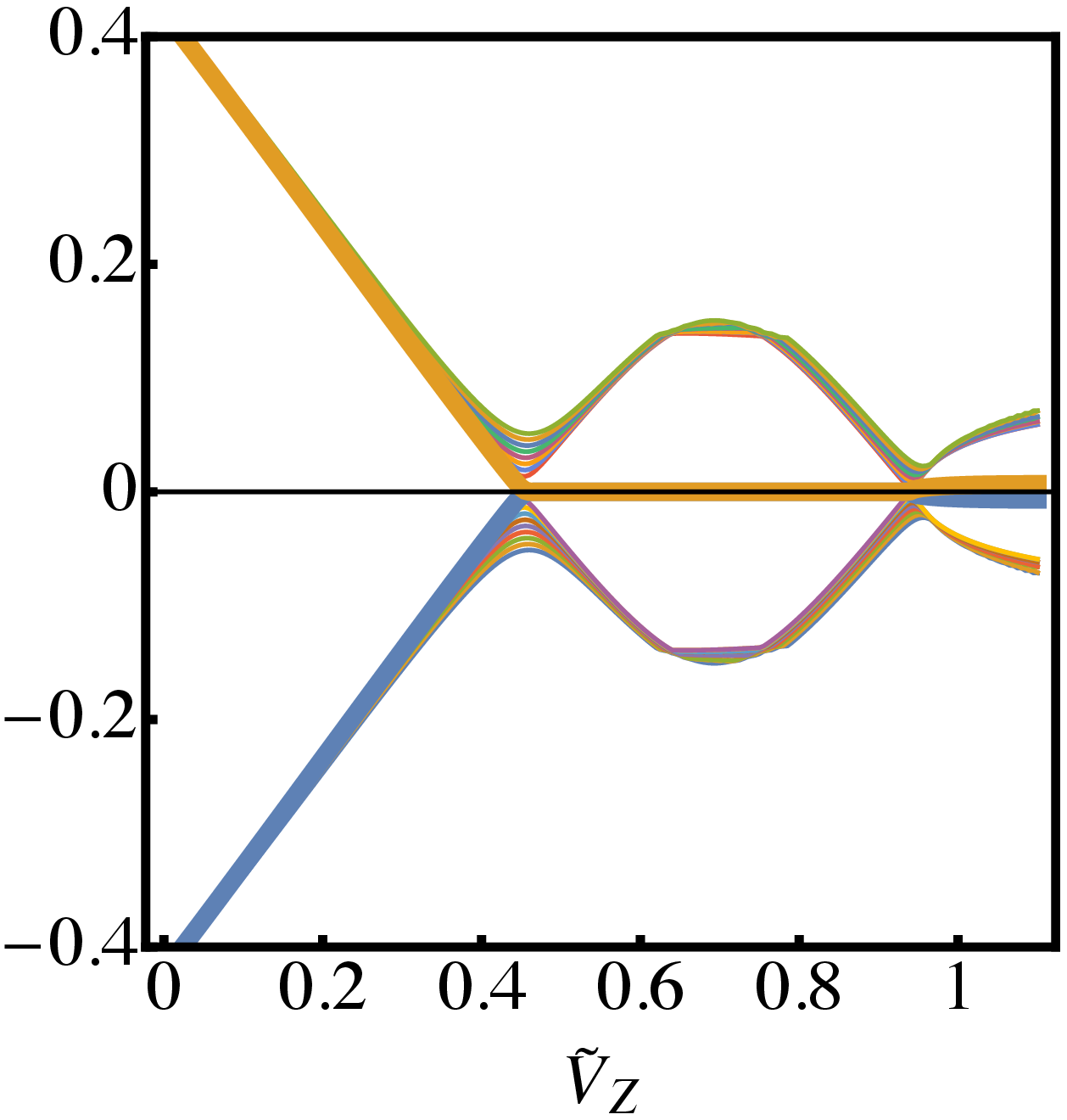}
\end{center}
\caption{The spectrum (the 20 lowest energy values) as a function of magnetic field, in the presence and in the absence of orbital effects (left and right panels respectively).  We take $\tilde{\Delta}=0.4$, $\tilde{\alpha}=0.2$,
 $\tilde{\mu}=3.9$, $g^*=50$.}  
\label{fig3}
\end{figure}

In addition, in Fig.~\ref{fig3} we plot the evolution of the spectrum with magnetic field when the chemical potential is fixed. Note how the energies of the lowest-lying four states merge to zero at some critical value $\tilde{V}_Z \approx 0.35$ marking the transition between the non-topological 
0 state and the topological state with two pairs of Majorana. For $\tilde{V}_Z$ larger than a second critical value $ \tilde{V}_Z  \approx 0.4$ two of these states acquire a finite energy and merge to the continuum marking the transition to the topological phase I exhibiting only the ``traditional'' pair of Majorana states.  In the absence of orbital field the state with two pairs of Majorana is absent and the two lowest-lying modes merge to zero and become Majorana for $\tilde{V}_Z$ larger than $\approx 0.4$.

\subsection{Discussion}
\begin{figure}[h]
\begin{center}
\includegraphics[width=.49\columnwidth]{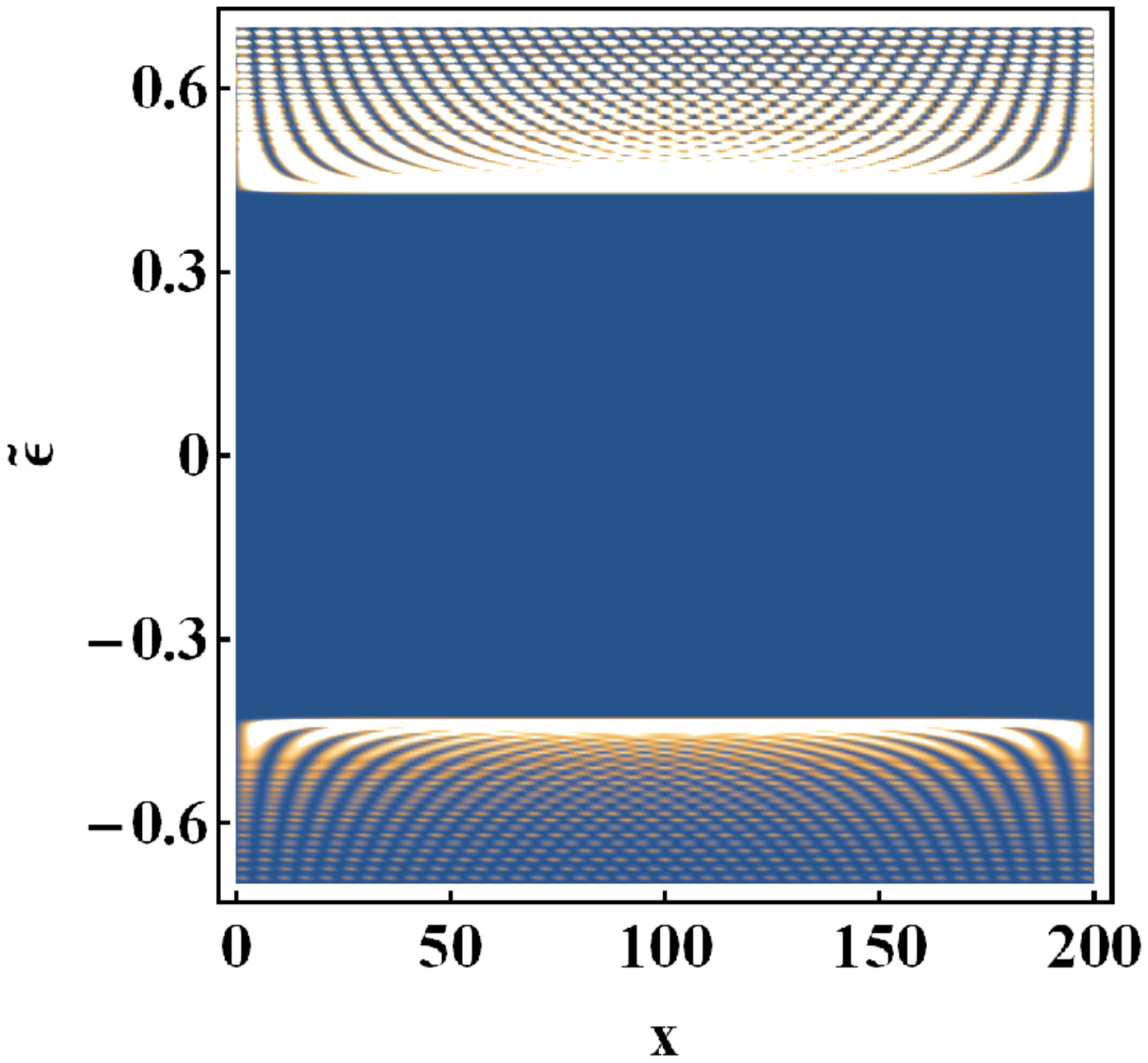}
\includegraphics[width=.49\columnwidth]{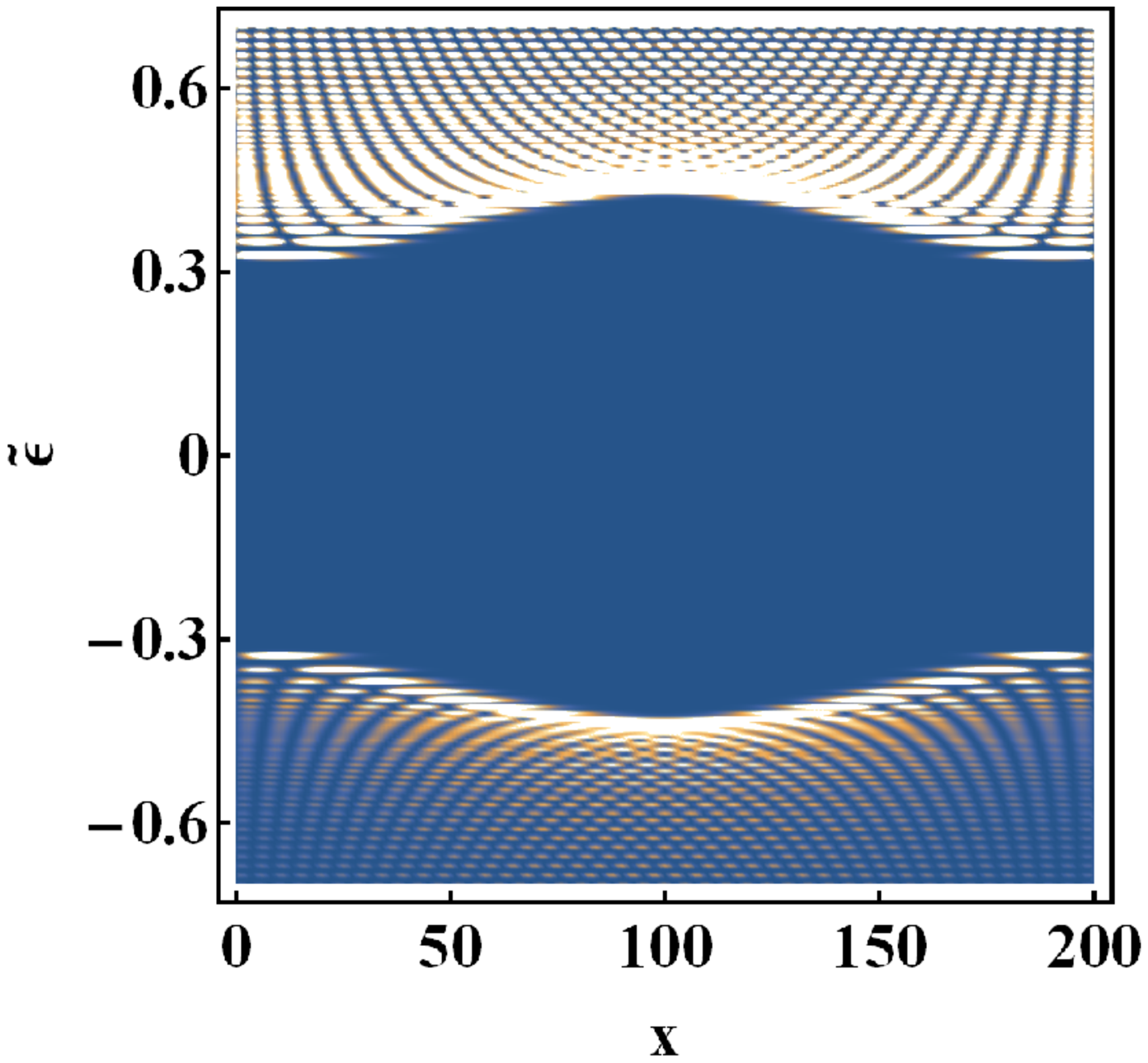}\\
\includegraphics[width=.49\columnwidth]{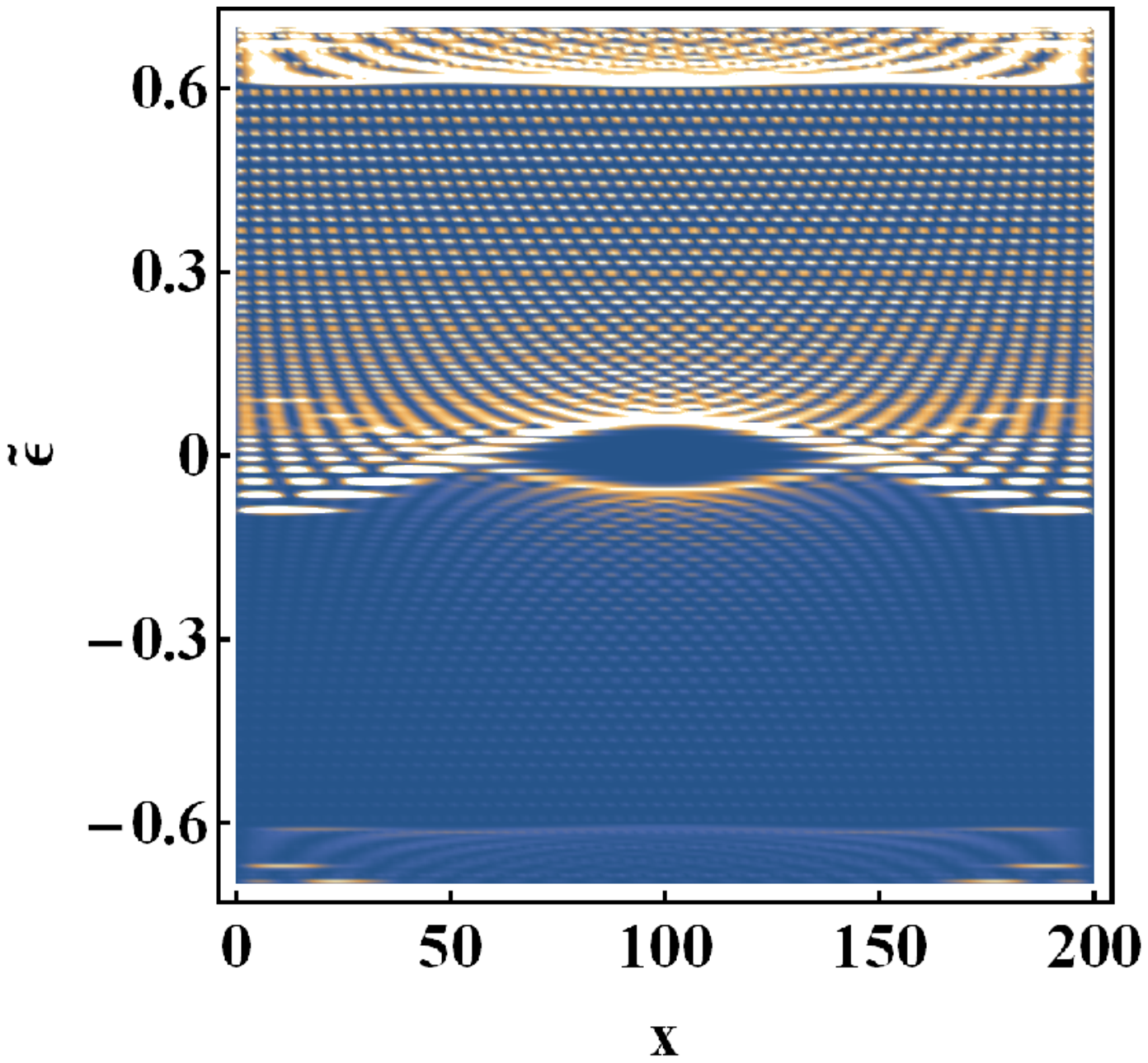}
\includegraphics[width=.49\columnwidth]{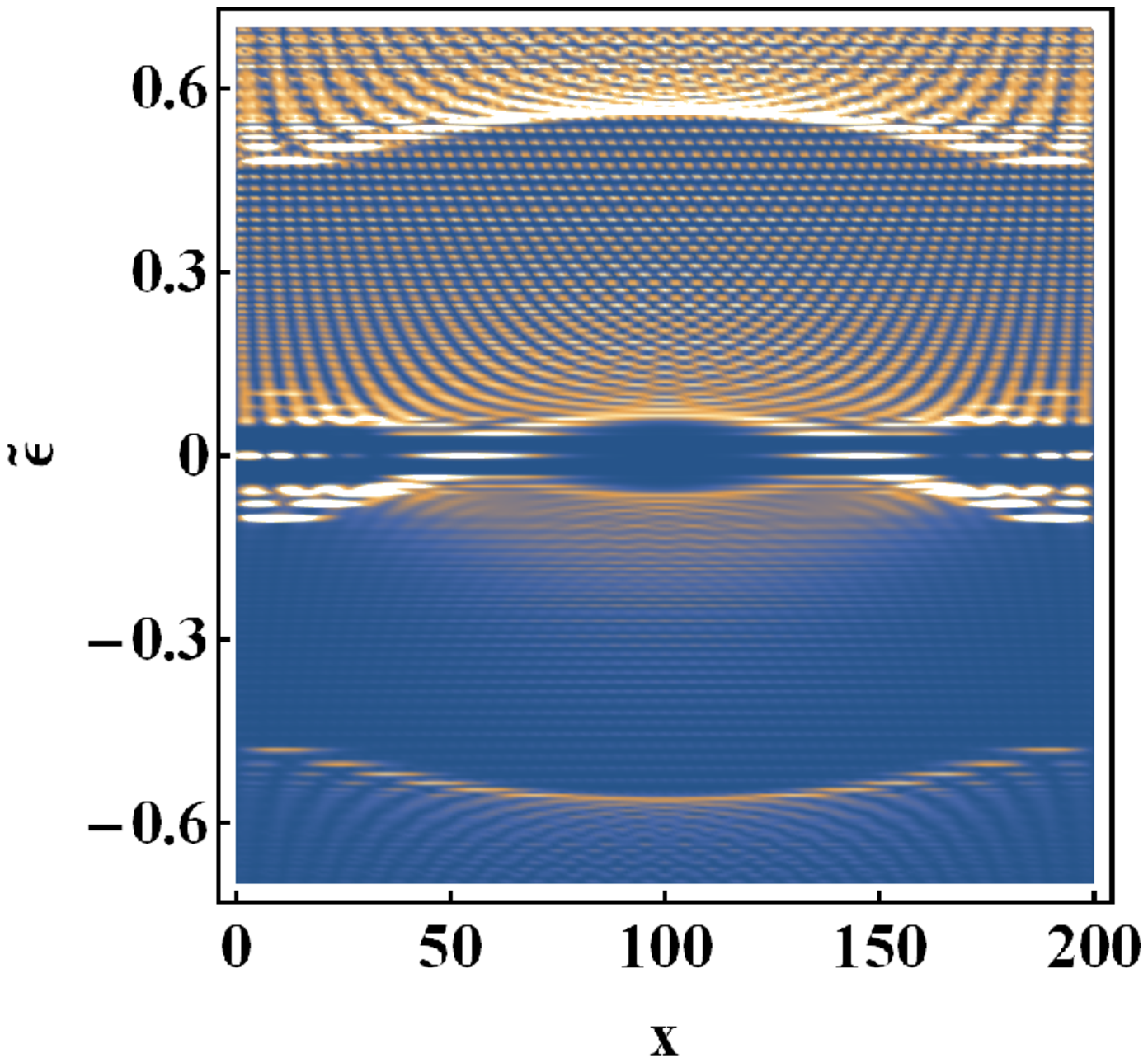}\\
\includegraphics[width=.49\columnwidth]{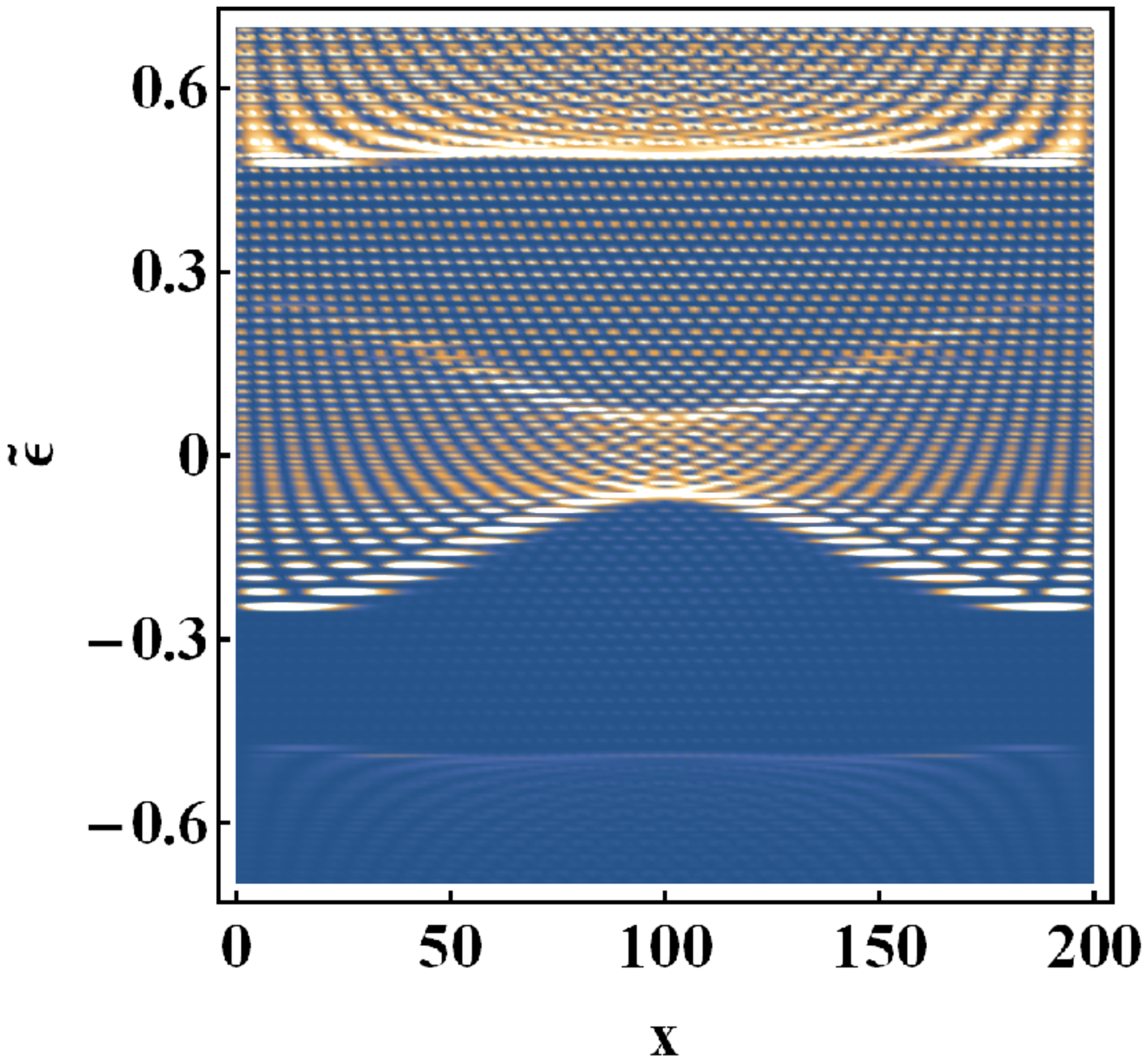}
\includegraphics[width=.49\columnwidth]{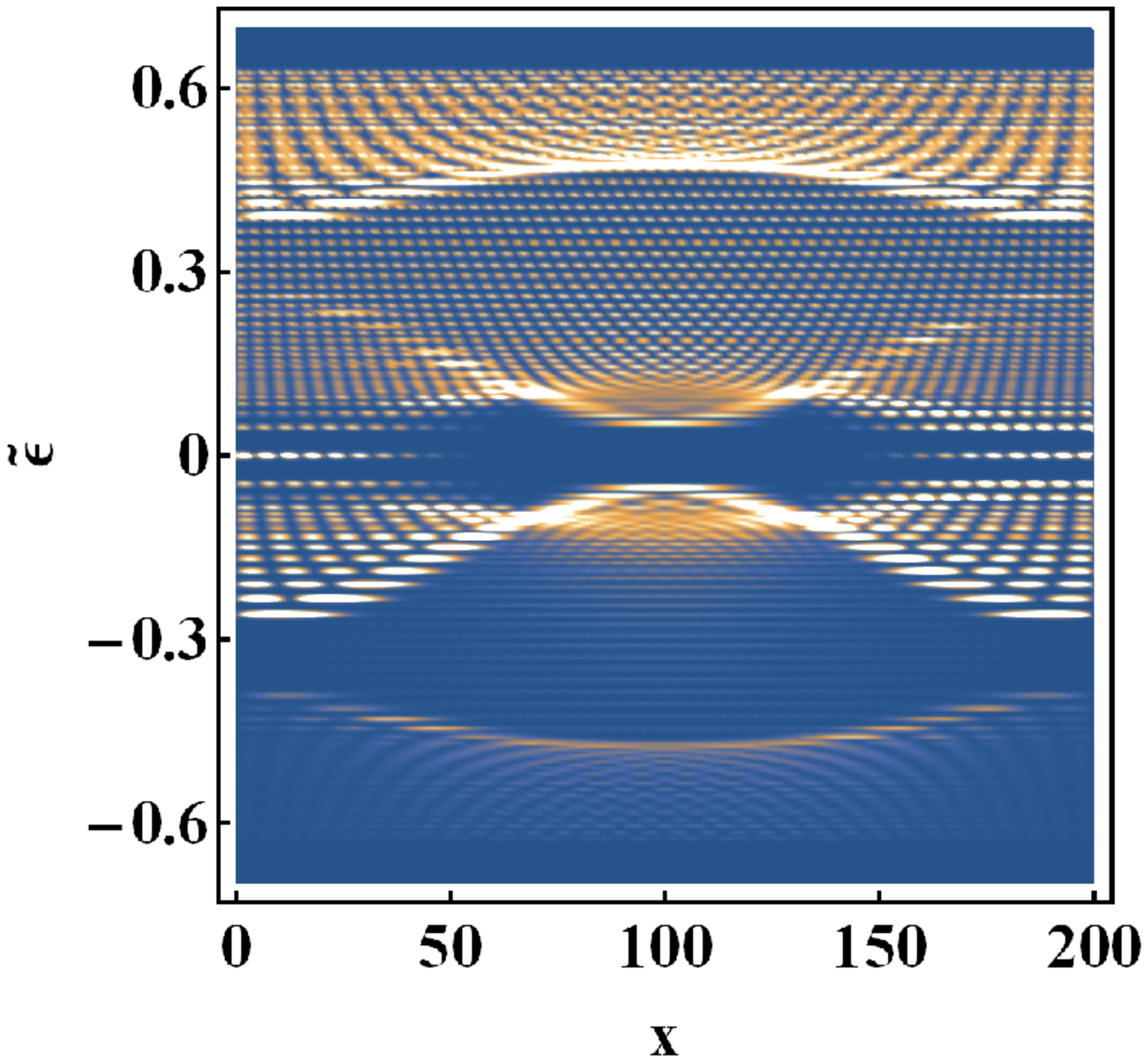}
\end{center}
\caption{The LDOS  (averaged along y) as a function of x and the energy when 
only SC is present, $\tilde{\Delta}=0.4$  (upper left panel), SC and orbital effects are present: $\tilde{\Delta}=0.4$, $\tilde{b}=0.3/50$ (upper right panel)
SC, orbital effects and Zeeman are present: $\tilde{\Delta}=0.4$, $\tilde{V}_Z=0.38$, $\tilde{b}=0.38/50$  (middle left panel), SC, orbital effects, Zeeman and spin-orbit are present:  $\tilde{\Delta}=0.4$, $\tilde{V}_Z=0.38$, $\tilde{b}=0.38/50$, $\tilde{\alpha}=0.2$ (middle right panel). For the lower two panels we take a larger magnetic field $\tilde{V}_Z=0.5$, $\tilde{b}=0.5/50$ and no spin-orbit (lower left panel) and $\tilde{\alpha}=0.2$ (lower right panel). }  
\label{dos}
\end{figure}

In order to understand better the formation of Majorana states we plot the DOS for the strip in the presence and absence of orbital effects as a function of energy and position. If only the SC is present, the system has a position-independent gap (slightly different from $\Delta$ because of the finite-size effects) (see the upper panel in the left column of Fig.~\ref{dos}).
If some orbital field is added (but no Zeeman field) we see that the uniform gap becomes effectively coordinate-dependent (Fig.~\ref{dos}, upper panel in the right column). 
When the actual Zeeman component is also taken into account along with the orbital field, the SC gap is reduced due to the Zeeman spin splitting. Given the spatial inhomogeneity of the effective gap in the presence of orbital effects, the gap closing caused by the Zeeman field will occur first in the regions close to the edges of the strip.  For a Zeeman field larger than a certain critical value this gives rise to two gapless regions separated by a gapped region in the bulk (see Fig.~\ref{dos}, middle panel - left column). 
When a non-zero spin-orbit coupling is added, these two regions become topological, giving rise each to a pair of Majorana states (Fig.~\ref{dos} middle panel - right column). This corresponds to phase II (red) in the topological phase diagram and the spatial profile of the two pairs of Majorana states is also depicted in Fig.~\ref{LDOSstrip}

When the Zeeman field becomes larger than a second critical value the entire strip becomes gapless (see Fig.~\ref{dos}, lower panel - left column).
When a spin-orbit is added only two Majorana states form, close to the ends of the wire, same as in the no-orbital case (Fig.~\ref{dos}, lower panel - right column), corresponding to the I region (white) in the topological phase diagram. 

We should note that even if the topological phase in which MBS can form is larger than in the absence of orbital effects, the topological gap protecting them is reduced, especially at high magnetic fields where the Majorana states are eventually destroyed. Also they are much more delocalized than in the case of no-orbital effects (see Appendix D for a plot of the LDOS in the presence of Majorana states when the orbital effects are not taken into account).

While not shown here, we have observed in our simulations that for longer wires the induced modifications of the gap by the orbital field may consist of multiple spatial oscillations. In the data presented here only two minima of the gap occur close to the edges of the wire, thus  one may obtain at most two pairs of MBS, however, depending on the size of the system and on the parameter values, more than two minima, and thus more that two topological regions may form in principle. 


\subsection{Square system}
\begin{figure}[h!]
\vspace{.2in}
\begin{center}
\includegraphics[width=0.49\columnwidth]{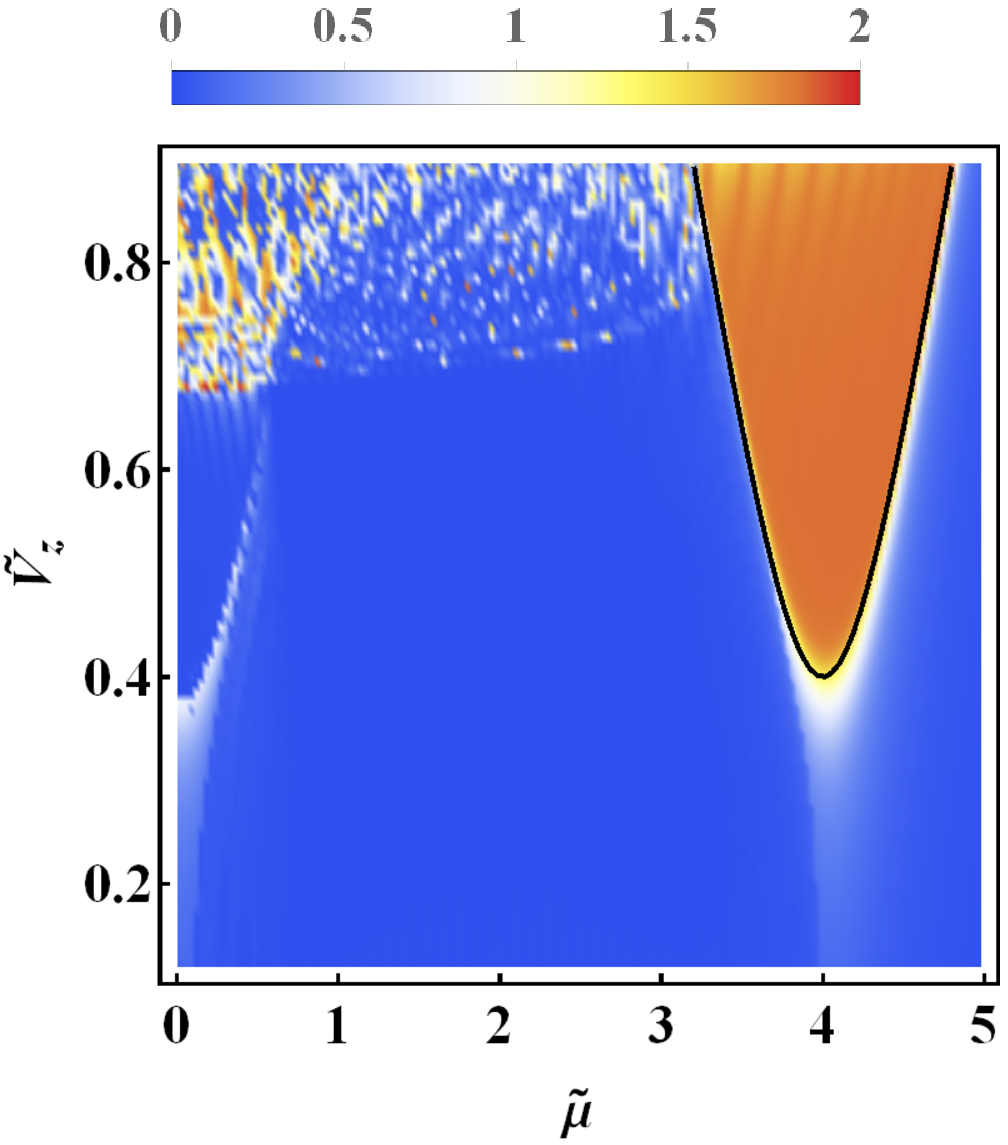}
\includegraphics[width=0.49\columnwidth]{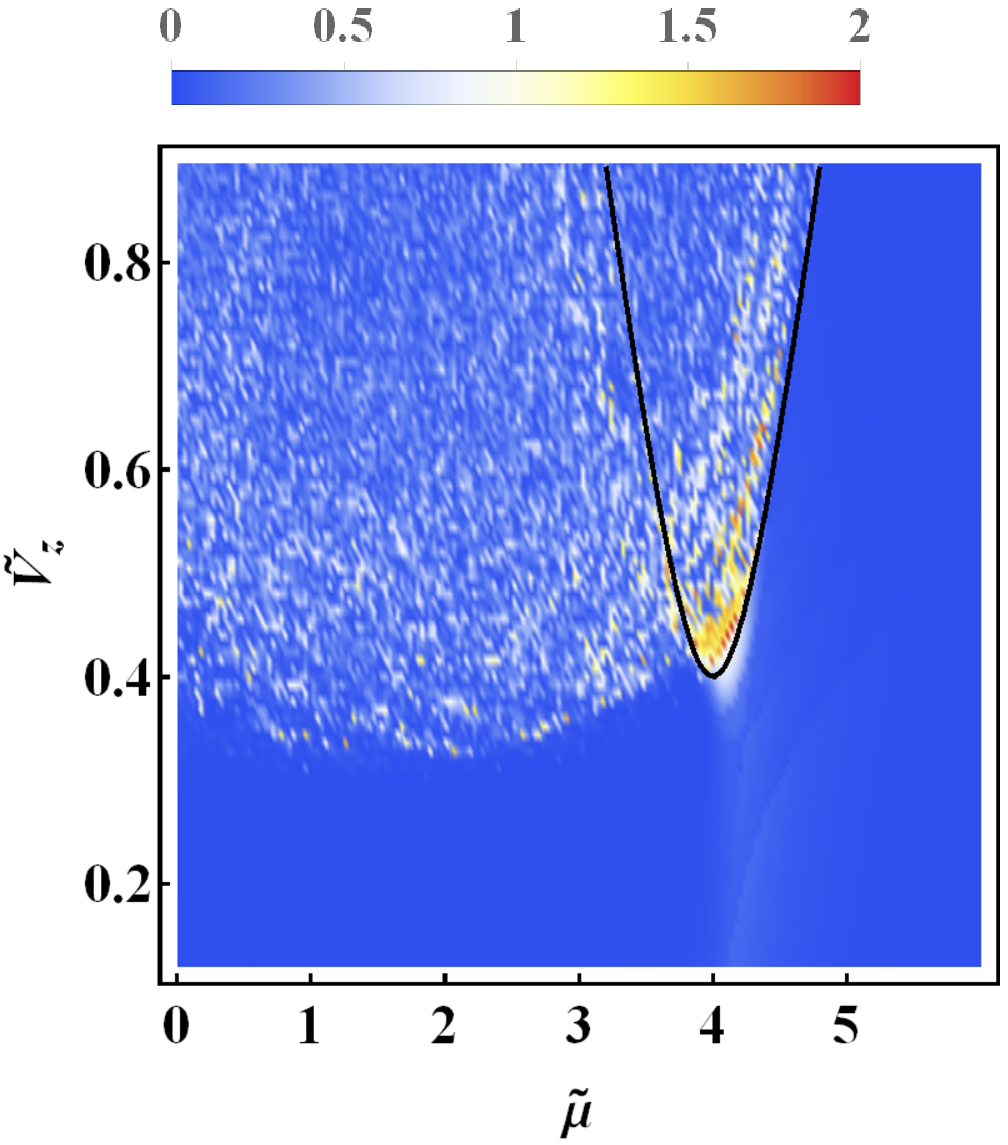}
\end{center}
\caption{Topological phase diagram as a function of $\tilde{\mu}$ and $\tilde{V}_Z$ for a square system. We take  $\tilde{\Delta}=0.4$, $\tilde{\alpha}=0.2$, $g^*=200$ (left panel), and $g^*=50$ (right panel).}  
\label{PhDsquares}
\end{figure}

In what follows, a square system of dimensions 51x51 is considered. In Fig.~\ref{PhDsquares}, the topological phase diagrams as functions of $\tilde{V}_Z$ and $\tilde{\mu}$ are presented for $g^*=50$. 
As depicted in Fig.~\ref{PhDsquares} we first note that, same as in the no-orbital-effects case, only quasi-MBS exist in this system, as predicted in Ref.~[\onlinecite{Sedlmayr2016}]. They are the result of the square shape of the system, which gives rise to overlaps between the wavefunctions that describe the MBS.  In the presence of orbital effects quasi-MBS persist only in the region close to $\tilde{\mu}=4$, and for low magnetic fields. A further increase of the orbital magnetic field destroys any structure of the topological phase diagrams. 

\section{Conclusions}
In this work, the orbital effects of the perpendicular component of the magnetic field have been studied for two planar systems with different geometries: a quasi-1D object (a strip) and a square system. 

In the case of a strip, one recovers MBS for extended regions of the topological phase diagram. These states persist even for relatively large values of the orbital magnetic field as long as $g m/m_e$ is relatively large. While the topological regions are more extended than in the absence of orbital effects, the topological gap protecting these states is significantly reduced.

Interestingly, more than one pair of Majorana bound states can be hosted by the strip. One of the pairs is the "traditional pair" localized at the edges of the strip, whereas the other one occurs closer to the middle of the strip. The presence of a second pair of MBS can be qualitatively explained by the fact that the orbital field introduces an effective coordinate dependence into the pairing term. Therefore, in the semiclassical picture the standard topological criterion takes a coordinate-dependent form $\sqrt{\mu'^2 + \Delta^2(x)} < V_Z$, which in-turn ensures that in certain parameter regimes the strip we consider consists of three regions: a topological superconductor on the left, a normal superconductor in the middle, and a topological superconductor on the right. Such a configuration supports 2 pairs of MBS, one for each of the topological regions. 

For the square system, due to the geometry of the problem, quasi-MBS appear as low-energy solutions of the tight-binding  Hamiltonian. We find that these quasi-MBS are less resistant to the effects of the orbital magnetic field.  




{\bf Acknowledgements}
VK would like to acknowledge interesting discussions with Alexander Zyuzin, Sergue\"i Tchoumakov and Egor Babaev. This work was supported by the Roland Gustafsson foundation for theoretical physics.

\bibliography{biblio_MForbital}

\widetext
\appendix   
\section{Mapping the Hamiltonian on the discrete grid}

The covariant gauge transformations along the $x$ direction are 
\begin{equation}
\Pi_x\psi(x,y) = \left( -i \hbar \partial_x + e A_x\tau_z\right) \psi(x,y)=e^{i \frac{e}{\hbar} b x y \tau_z} \left(-i \hbar \partial_x \right) e^{-i \frac{e}{\hbar} b x y\tau_z } \psi(x,y)
\label{A1}
\end{equation}
 for the Landau gauge and
\begin{equation}
\Pi_x\psi(x,y) = \left( -i \hbar \partial_x + e A_x\tau_z \right) \psi(x,y) = e^{i \frac{e}{2\hbar} b x y \tau_z} \left(-i \hbar \partial_x \right) e^{-i \frac{e}{2\hbar} b x y\tau_z } \psi(x,y)
\label{A2}
\end{equation}
for the symmetric gauge. On the discrete grid, these map to
\begin{equation}
\Pi_x\psi(x,y)= -\frac{i \hbar}{2a}\Big[e^{-i \frac{e}{\hbar} b a y\tau_z } \psi(x+a,y)-e^{i \frac{e}{\hbar} b a y\tau_z } \psi(x-a,y)\Big]
\label{A3}
\end{equation}
and 
\begin{equation}
\Pi_x\psi(x,y)= -\frac{i \hbar}{2a}\Big[ e^{-i \frac{e}{2\hbar} b a y\tau_z } \psi(x+a,y)-e^{i \frac{e}{2\hbar} b a y\tau_z } \psi(x-a,y)\Big]
\label{A4}
\end{equation}
correspondingly. Above we have used a standard procedure for discretizing derivatives.
Translating this into the tight-binding language implies that the terms containing the $x$-components of the momentum operator, like the spin-orbit coupling will acquire a phase factor $e^{-i \frac{e}{\hbar} b a y\tau_z }$ for the Landau gauge and $e^{-i \frac{e}{2\hbar} b a y\tau_z }$ for the symmetric gauge. The kinetic term (corresponding here to the hopping term) contains a second-order derivative and  behaves as follows
\begin{equation}
\Pi^2_x\psi(x,y) = -\frac{\hbar^2}{2a^2} \left[  e^{-i \frac{e}{\hbar} b a y\tau_z } \psi(x+a,y)+e^{i \frac{e}{\hbar} b a y\tau_z } \psi(x-a,y)-2 \psi(x,y) \right]
\label{A5}
\end{equation}
\begin{equation}
\Pi^2_x\psi(x,y) = -\frac{\hbar^2}{2a^2} \left[  e^{-i \frac{e}{2\hbar} b a y\tau_z } \psi(x+a,y)+e^{i \frac{e}{2\hbar} b a y\tau_z } \psi(x-a,y)-2 \psi(x,y) \right]
\label{A6}
\end{equation}
for the Landau and symmetric gauges respectively. This implies that the hopping terms in the tight-binding Hamiltonian will acquire the same phase factors as the spin-orbit terms. Along the same lines we can discretize $\Pi_y \psi(x,y)$ and $\Pi^2_y \psi(x,y)$. Thus in the symmetric gauge the terms containing the $y$-components of the momentum will acquire the phase factors $e^{\mp i \frac{e}{2\hbar} b a x\tau_z }$  instead of $e^{\pm i \frac{e}{2\hbar} b a y\tau_z }$ respectively, whereas in the Landau gauge no phase factors will appear since $A_y = 0$.  Eqs.~(\ref{A1}-\ref{A6}) allows to conclude that the effects of the orbital field are captured in the tight-binding form of the Hamiltonian by the introduction of a phase factor $e^{-i \frac{e}{\hbar} b a y\tau_z }$ in the Landau gauge versus $e^{-i \frac{e}{2\hbar} b a (-x + y)\tau_z }$ in the symmetric gauge for both the hopping and the spin-orbit coupling terms. 

The superconducting pairing term has to be modified as well. The wavefunction is changed in the following manner:
\begin{equation}
\Psi'(x,y)=e^{-i \frac{e}{\hbar} b x y\tau_z } \Psi(x,y)
\end{equation}
and the gap should be modified as follows: $\Delta'=e^{-2 i \frac{e}{\hbar} b x y\tau_z } \Delta$. The numerical analysis shows, however, that one needs to use rather $\Delta'=e^{-i \frac{e}{\hbar} b x y\tau_z } \Delta$ to preserve gauge invariance in the Landau gauge. The redundant factor of $2$ is a drawback of using the mean-field approximation to the BCS theory. The correct expression can be derived using a self-consistent gap calculation\cite{Arseev2006}. In the symmetric gauge, on the other hand, no phase factors appear for the superconducting pairing term.

\section{Hoftsatder butterfly}
\begin{figure}[h!]
\begin{center}
\includegraphics[width=0.4\columnwidth]{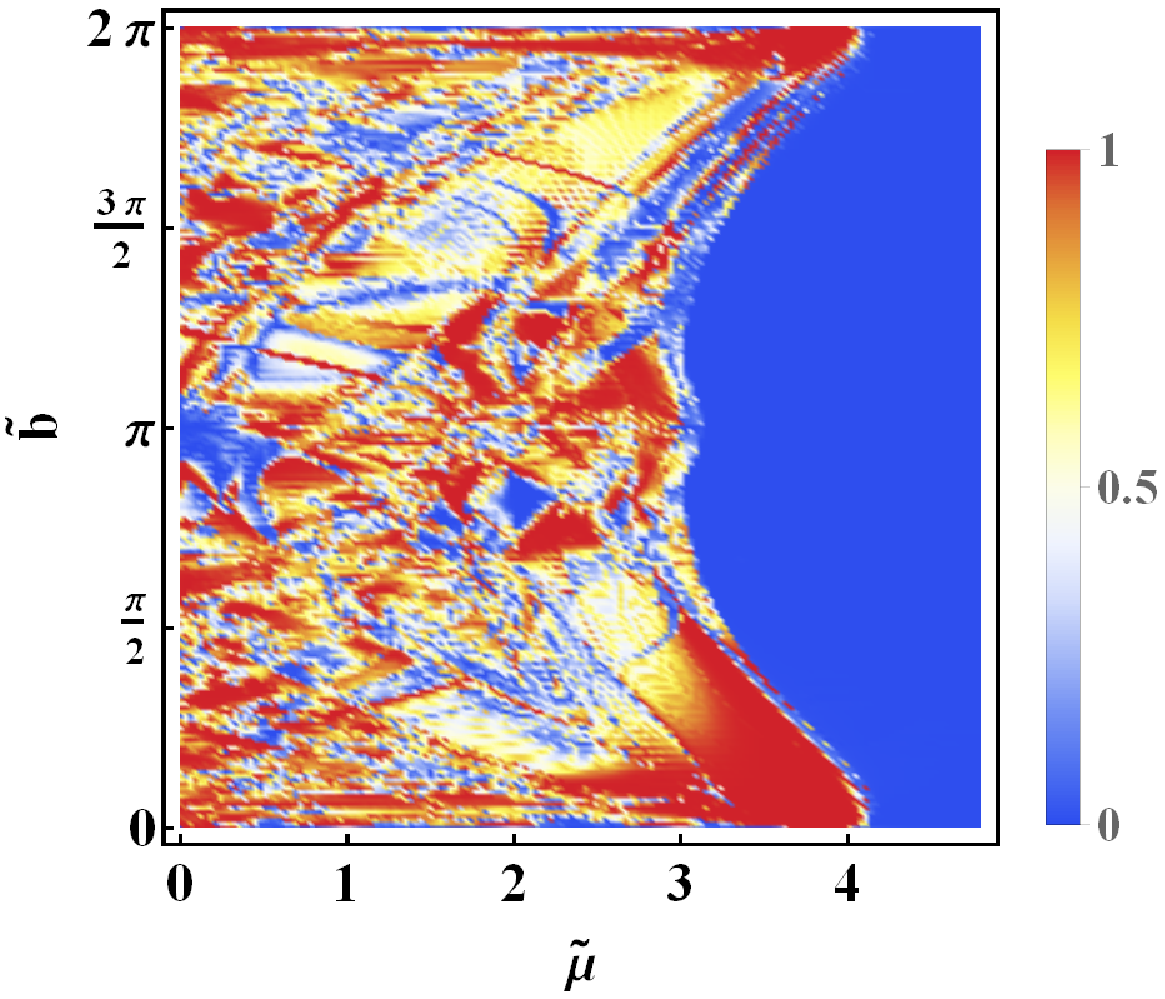}
\includegraphics[width=0.4\columnwidth]{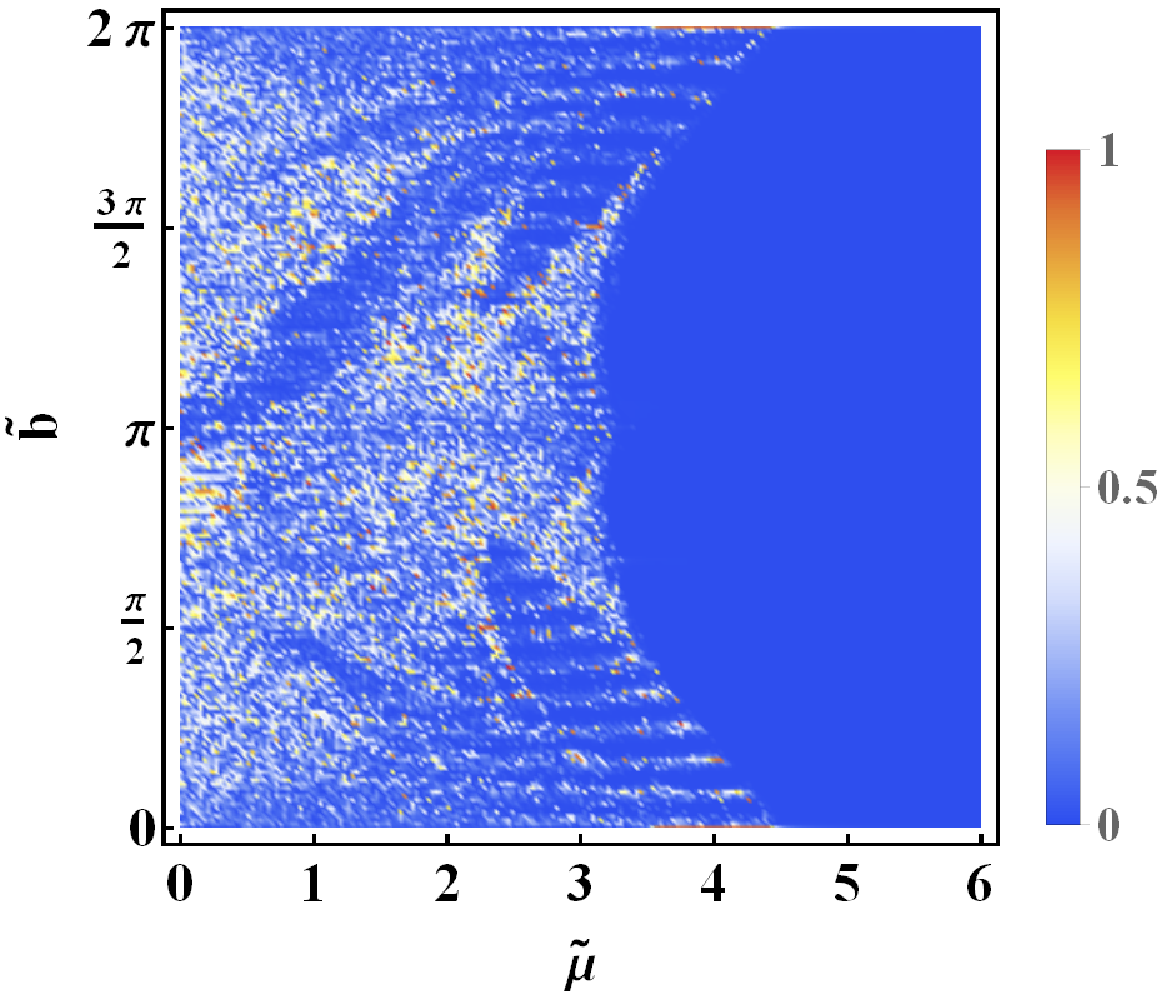}
\end{center}
\caption{The topological phase diagrams as a functions of $\tilde{b}$ and $\tilde{\mu}$ with a fixed Zeeman field. We set $\tilde{V}_Z=0.5$ and $ \tilde{V}_Z=0.6$ for the left and right panels respectively. The size of the system is $5 \times 201$ for the left  panel and $51 \times 51$ for the right one. Other parameters: $\tilde{\Delta}=0.4; \tilde{\alpha}=0.2$.} 
\label{Hofstadter}
\end{figure}

In Fig.~\ref{Hofstadter} we plot the  MP as a function of both $\tilde{\mu}$ and $\tilde{b}$ assuming that $V_Z$ is fixed (not necessarily physical but useful to understand the effects of the magnetic field when these effects are very large). The resulting structure reminds us of the Hofstadter butterfly, a fractal structure that appears in the spectrum of 2D systems when an orbital magnetic field is introduced.

\section{Special point $\bs{\tilde{b}=\pi}$}

For $\tilde{b}=\pi$, one notices how the phase diagram regains some structure and has some similarities to the topological phase diagram for a one-dimensional system (see e.g. Ref.~[\onlinecite{Sedlmayr2015a}]).
 \begin{figure}[h!]
\begin{center}
\includegraphics[width=0.35\columnwidth]{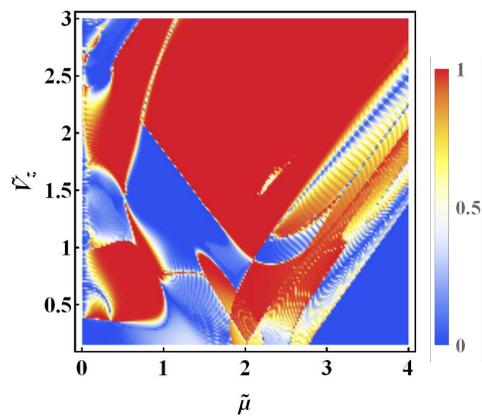}
\end{center}
\caption{The topological phase diagram $\tilde{V}_Z$-$\tilde{\mu}$ for a 5 $\times$ 201 system for $\tilde{b}=\pi$. Other parameters: $\tilde{\Delta}=0.8; \tilde{\alpha}=0.4$.}
\label{pi}
\end{figure}

\section{Formation of MBS without orbital field}

\begin{figure}[h]
\begin{center}
\includegraphics[width=.35\columnwidth]{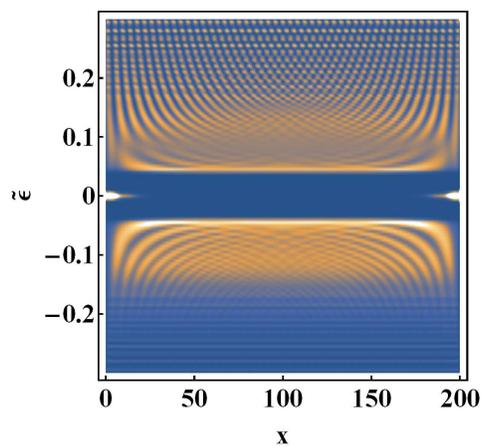}
\end{center}
\caption{The density of states plotted as a function of position $x$ and dimensionless energy $\tilde{\epsilon}$. The bright spots at zero energy represent the ``traditional pair'' of MBS localized at the ends of the system in the absence of orbital effects. We set $\tilde{\Delta}=0.4$, $\tilde{V}_Z=0.5$,  $\tilde{\alpha}=0.2$.}  
\label{dos2}
\end{figure}

\end{document}